\documentclass[final]{aipproc}

\layoutstyle{6x9}

\usepackage[utf8x]{inputenc}
\usepackage[english]{babel}
\usepackage{amssymb}
\usepackage{amsopn}
\usepackage{amsmath} 

\usepackage{graphicx}
\begin{document}

\title{Lipatov's Action Beyond Tree Level}


\author{José Daniel Madrigal Martínez}{
  address={Instituto de Física Teórica UAM/CSIC, Nicolás Cabrera 15\\ C.U. Cantoblanco, E-28049 Madrid (Spain)}
}

\begin{abstract}
Computations in Quasi-Multi-Regge kinematics are much simplified using an effective theory pioneered by Lipatov. This high-energy effective theory does not come from an integrating out procedure, and suitable counterterms and regularization have to be provided beyond tree level. In this note we explain this prescription and illustrate it with one- and two-loop examples. 
\end{abstract}

\maketitle

\section{The High-Energy Effective Action}

Notable simplifications arise in QCD scattering amplitudes in the multi-Regge limit, where center-of-mass energy $s_i$ is much bigger than the momentum transfer $t_i$ for each subchannel in a particle production process. In this high-energy region, QCD becomes very similar to its supersymmetric counterpart, ${\cal N}=4$ SYM, and two-dimensional conformal invariance, with links to integrability, appears \cite{Lip1986,Lip1993,FK1995}. Moreover, the property of reggeization allows to resum the dominant contributions of the form $[\alpha_s\ln(s_i/t_i)]^n$ (and $\alpha_s[\alpha_s\ln(s_i/t_i)]^n$, to NLO) within the BFKL approach \cite{FKL1977,BL1978,IFL2010}. The defining hierarchy of scales of the high-energy limit calls for an effective theory formulation, where the enhanced symmetry properties become more manifest. Moreover, the effective action is the most direct way to provide all the necessary elements to build all the contributions required by unitarity, that are in general missed by linear BFKL evolution and are only partially built in other approaches like the color glass condensate (CGC) \cite{Tri2005}.\\

\begin{figure}[htp]
\centering
\includegraphics[scale=0.5]{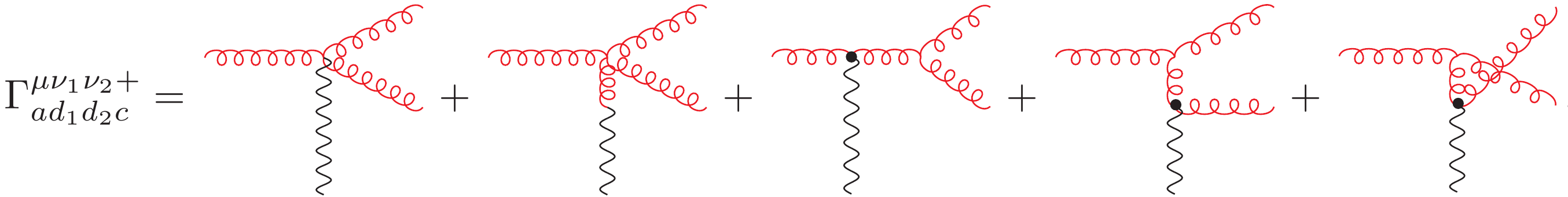}
\caption{An infinite number of induced vertices are needed to preserve the gauge invariance of the reggeon--$N$-gluon vertex, for all values of $N$.}
\label{1fig}
\end{figure}

The effective action proposed by Lipatov in \cite{Lip1995,Lip1997} reads\footnote{The following Sudakov decomposition applies: $k=\frac{1}{2}(k^+n^-+k^-n^+)+{\bf k})$, $n^{+,-}=2p_{A,B}/\sqrt{s}$, where $p_{A,B}$ are the momenta of the initial particles.}
\begin{equation}\label{1}
\begin{aligned}
&S_{\rm eff}=S_{\rm QCD}+S_{\rm ind};\quad S_{\rm ind}=\int d^4x\,{\rm Tr}\left[(W_+[v(x)]-A_+(x))\partial_\perp^2 A_-(x)\right]+\{+\leftrightarrow -\},\\
&W_\pm[v(x)]=-\frac{1}{g}\partial_\pm{\cal P}\exp\left(-\frac{g}{2}\int_{-\infty}^{x^\mp}dz^\pm v_\pm(z)\right)=v_\pm-gv_\pm\frac{1}{\partial_\pm}v_\pm+\cdots
\end{aligned}
\end{equation}
where $A_\pm$ is the $t$-channel reggeon field, which interacts \emph{locally in rapidity} with gluon fields $v_\mu$ by means of Wilson lines $W_\pm(v)$. Projection on multi-Regge kinematics imposes the constraints $\partial_\pm A_\mp=0$. In the original construction, the induced terms in $S_{\rm ind}$ appeared by requiring gauge invariance of the effective vertices for quasielastic $N$-gluon production. At each order $N$, a new induced term has to be added to preserve this property (\textsc{Fig.} \ref{1fig}). Ward identities can be then derived in the form of recurrence relations, which in particular allow for a general expression for the Feynman rule of the $N$-gluon--reggeon effective vertex \cite{ALKC2005}. The perturbative expansion can then be cast in the form of Wilson line \eqref{1}, something that can be expected on general grounds for the 2d effective high-energy theory of QCD \cite{VV1993}. Wilson lines appear as well, following a different route, in other formalisms of high-energy QCD like the CGC \cite{GIJV2010,JKMW1997,Hat2007} or the high-energy operator product expansion \cite{Bal1996}. Integrating out the physical degrees of freedom in \eqref{1} Gribov's reggeon field theory \cite{Gri1968} should be recovered \cite{Lip1997}.
\begin{figure}[htp]
\centering
\includegraphics[scale=0.5]{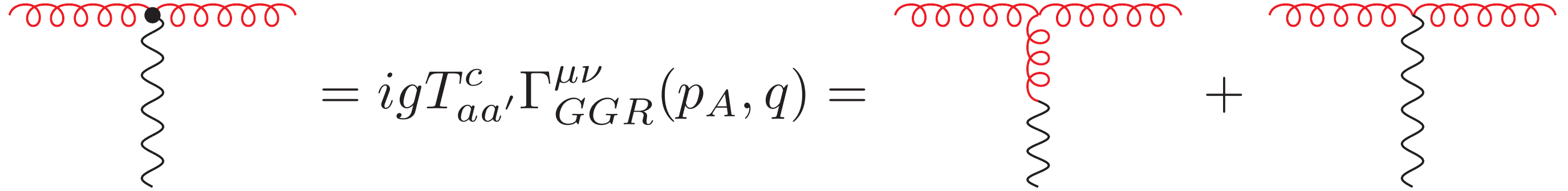}
\end{figure}
\section{Lipatov's Action and Loops}

\begin{figure}[htp]
\centering
\includegraphics[scale=0.6]{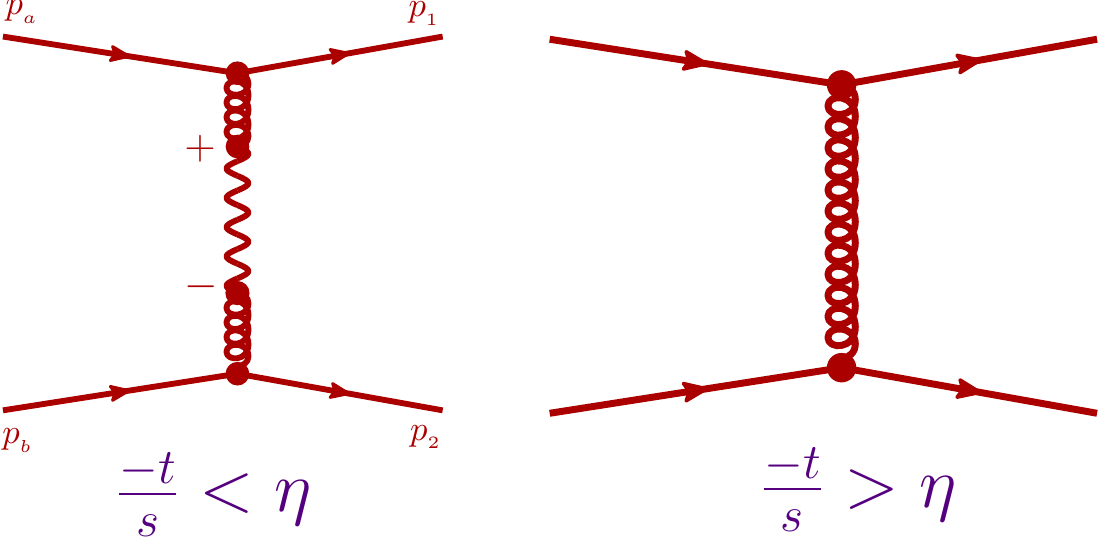}
\caption{Both diagrams give the same result in the high-energy limit, but only the first one should be considered when the interacted particles are far away in rapidity, avoiding overcounting.}
\label{2fig}
\end{figure}

At leading log accuracy, where no final particles are allowed to be close in rapidity, a formulation of the effective action in the Wilsonian sense of integrating out highly virtual modes exist \cite{Lip1991,KLS1994,KNS1995}. However, no such formulation of the effective action for processes beyond multi-Regge kinematics is known at present. Instead, a new term is added in \eqref{1} to the \emph{full} QCD action. The apparent overcounting is cured by remembering that $S_{\rm ind}$ describes interactions with particles in a limited range of rapidities, and in order to preserve quasi-multi-Regge factorization clusters of particles with very different rapidities can only be connected through reggeon exchange. This requirement can be enforced by an explicit cutoff (\textsc{Fig.} \ref{2fig}), but this becomes really unwieldy when loops appear. Another difficulty arising beyond tree level is the appearance of a new set of divergences related to the operators $\partial_\pm^{-1}$ in \eqref{1}, as $k^\pm\to 0$.\\

\begin{figure}[htp]
\centering
\includegraphics[scale=1]{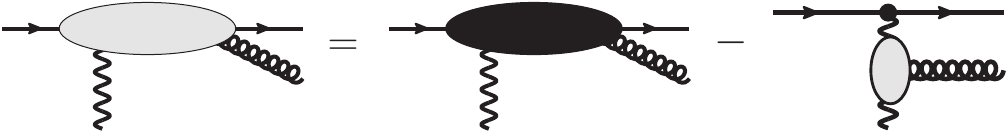}
\caption{Subtraction of nonlocal contributions in the real corrections to the forward jet vertex. The subtraction corrects for the integration region in the loop for which the outcoming gluon carries almost all the initial quark momentum, giving rise to a large rapidity separation between the final particles.}
\label{3fig}
\end{figure}

In \cite{HV2012,CHMV2012} (see \cite{CHMV2012b} for a recent review) a procedure was devised to deal with these problems. In the first place, a gauge-invariant regularization was given by introducing the tilted light-cone vectors $n^{a,b}=n^{+,-}+e^{-\rho}n^{-,+}$. The regulator $\rho$ can be identified formally as $\ln s$. In the end, one should take the high-energy limit $\rho\to\infty$ and physical results must be independent of the regulator $\rho$. On the other hand, the cutoff procedure is substituted by the subtraction of those diagrams involving nonlocal reggeon exchange, as exemplified in \textsc{Fig.} \ref{3fig}. This procedure can be rephrased as a renormalization of the reggeon wavefunction and the effective vertices \cite{CHMV2012,CHMV2012b}; in this context the subtractions can be thought as relevant counterterms for the reggeon action.


\section{Forward Jet Vertex and Gluon Regge Trajectory}

The subtraction-regularization prescription has been shown in \cite{HV2012,CHMV2012} to reproduce the results obtained through unitarity techniques for the 1-loop quark-initiated forward jet vertex \cite{BCV2002,CIMPP2012} and the quark piece of the 2-loop gluon trajectory \cite{FFK1996a}. Recently, the gluon-initiated forward jet vertex has also been derived in this formalism \cite{CHMVp}. Convolution of this
jet vertex with the NLO BFKL gluon Green function is expected to play a very important role in the description of
jet production at the LHC physics program.\\

\begin{figure}[htp]
\centering
\includegraphics[scale=0.75]{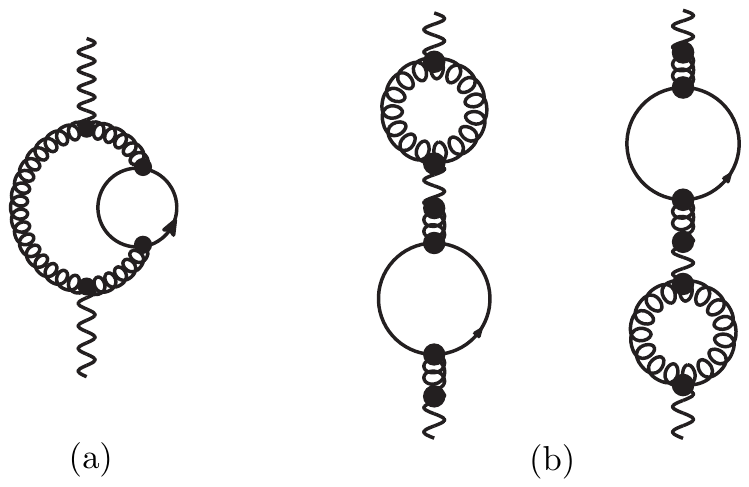}
\caption{Relevant diagrams in the high-energy limit for the computation of the piece of the gluon Regge trajectory proportional to $N_f$; subtractions are shown in (b).}
\label{4fig}
\end{figure}

An important advantage of the subtraction mechanism is that the powerful techniques in the literature can be applied to the computation of the appearing loop integrals (see, e.g. \cite{Smi2006}). Moreover, it captures subleading terms in the regulator $\rho$ that would not be taken into account by a naive cutoff \cite{CHMV2012b}. The number of relevant diagrams is also heavily reduced. In the computation of quark loops for the 2-loop gluon trajectory, actually only two independent diagrams proportional to $\rho$ (\textsc{Fig.} \ref{4fig}) must be computed\footnote{In general, for the computation of the trajectory, involving the self-energy corrections to the reggeon propagator, scaling arguments can be given to show that only those diagrams where the reggeon couples on both ends to more than one reggeon can be enhanced in $\rho$ and hence needs to be computed.}.\\

Having passed a number of nontrivial checks, we expect now to apply this computational technique to other processes relevant for high-energy phenomenology, like jet-gap-jet events. It would be also interesting to obtain an independent computation of the NLO photon impact factor \cite{BC2010} with this technique. Progress is also needed in mathematically formalizing this program.
\begin{theacknowledgments}
I am grateful to G. Chachamis, M. Hentschinski and A. Sabio Vera, with whom this work was carried out. This research is supported by E. Comission [LHCPhenonet (PITN-GA-2010-264564)] and C. Madrid [HEPHACOS ESP-1473]
\end{theacknowledgments}

\bibliographystyle{aipproc}   
\bibliography{Books.bib}{}

\end{document}